\documentclass[reprint, prmaterials, superscriptaddress, aps, amsmath, amssymb]{revtex4-2}

\usepackage{siunitx}
\usepackage{graphicx}
\usepackage[T1]{fontenc}
\DeclareSIUnit{\angstrom}{\textup{\AA}}
\date{August 26, 2026}

\begin{document}

\title{Interlayer-engineering of Charge Order Wave Vector in Kagome Metals}

\author{Muntafa M. Mahi}
\affiliation{Department of Electrical and Electronic Engineering, Bangladesh University of Engineering and Technology, Dhaka 1000, Bangladesh}

\author{Quazi D. M. Khosru}
\affiliation{Department of Electrical and Electronic Engineering, Bangladesh University of Engineering and Technology, Dhaka 1000, Bangladesh}
\author{M. Zahid Hasan}
\affiliation{Laboratory for Topological Quantum Matter and Advanced
Spectroscopy, Department of Physics, Princeton University, Princeton,
New Jersey 08544, USA}

\author{Mahbub Alam}
\affiliation{Department of Electrical and Electronic Engineering, Bangladesh University of Engineering and Technology, Dhaka 1000, Bangladesh}
\author{Md Shafayat Hossain}
\email{Contact Author: shossain@seas.ucla.edu}
\affiliation{Department of Materials Science and Engineering, University of
California, Los Angeles, California 90095, USA}
\affiliation{California NanoSystems Institute, University of California, Los Angeles, California 90095, USA}

\begin{abstract}
Charge orders in the kagome metals \textit{A}V$_3$Sb$_5$ sit at the center of a rich phase diagram that also includes superconductivity, nematicity, and signatures of time-reversal-symmetry breaking. Yet even the basic question of which charge ordering wave vectors are intrinsic, and which are selected by dimensionality and lattice coupling, remains unsettled. Importantly, the microscopic origin of different charge orders and, in particular, the relationship between the robust bulk $2\times 2$ charge order and the controversial $4\times 1$ modulation, which is primarily resolved by surface probes, remains unresolved. Here, we use first-principles calculations to study the role of interlayer coupling in CsV$_3$Sb$_5$ by tuning the interlayer separation from the monolayer limit to the bulk limit. In the monolayer \textit{A}V$_3$Sb$_5$ (\textit{A} = Rb, Cs), the phonon spectrum exhibits no instability at the M point; instead, the dominant lattice instability occurs at $\text{q} = (1/4, 0, 0)$, consistent with a $4\times 1$ modulation. As interlayer coupling increases in CsV$_3$Sb$_5$, an M-point phonon progressively softens and becomes unstable already near $c\approx \qty{12.24}{\angstrom}$, evolving into the strong $2\times2$ instability characteristic of the bulk. These results identify interlayer coupling as a control parameter at a fixed stoichiometry that links competing $4\times 1$ and $2 \times 2$ tendencies, providing a unified framework for understanding why multiple charge-order wave vectors coexist and compete in kagome metals.
\end{abstract}
\maketitle

In a metal, the Fermi surface is not merely a passive boundary in momentum space. It encodes the susceptibility to symmetry breaking~\cite{scalapinoCommonThreadPairing2012,wuNatureUnconventionalPairing2021,johannesFermiSurfaceNesting2008}. When the electronic density of states is enhanced, when scattering is structured in momentum, or when the lattice couples strongly and selectively to electronic states, the system can reorganize into a new collective phase~\cite{varmaStrongCouplingTheoryChargeDensityWave1983, wangCompetingElectronicOrders2013, kieselUnconventionalFermiSurface2013,kieselSublatticeInterferenceKagome2012}. Charge orders are the canonical examples of this phenomenon, and they also set the stage for a rich phase diagram consisting of multiple, and sometimes competing quantum phases and Fermi surface instabilities, including intertwined order, dimensional crossover, and unconventional superconductivity~\cite{fradkinColloquiumTheoryIntertwined2015,calandraEffectDimensionalityChargedensity2009, morosanSuperconductivityCuxTiSe22006}. The kagome metals \textit{A}V$_3$Sb$_5$ (\textit{A} = K, Rb, Cs) exemplify such a scenario. Their kagome-derived electronic structure hosts van Hove singularities (vHS) near the Fermi level, strong momentum-selective scattering, and a charge order transition at temperatures well above the onset of superconductivity~\cite{jiangUnconventionalChiralCharge2021, kangTwofoldVanHove2022,huRichNatureVan2022,liUnidirectionalCoherentQuasiparticles2023}. Experiments further indicate that the charge order phase is accompanied by additional symmetry breaking, including nematicity and optical signatures that have been interpreted as time-reversal-symmetry breaking~\cite{xingOpticalManipulationChargedensitywave2024,nieChargedensitywavedrivenElectronicNematicity2022,xuThreestateNematicityMagnetooptical2022,wuUnidirectionalElectronPhonon2023}. The result is a phase diagram where charge order, superconductivity, and symmetry breaking appear highly entangled, making \textit{A}V$_3$Sb$_5$ a model platform for the broader question of how electronic and lattice degrees of freedom conspire to select emergent phases~\cite{neupertChargeOrderSuperconductivity2022}.

Despite intense attention, the microscopic origin of the charge order in \textit{A}V$_3$Sb$_5$ remains unsettled~\cite{dennerAnalysisChargeOrder2021,christensenTheoryChargeDensity2021,ortizFermiSurfaceMapping2021,wuChargeDensityWave2022,wangOriginChargeDensity2022}. In the bulk, diffraction and related probes establish a dominant $2\times2$ reconstruction~\cite{liObservationUnconventionalCharge2021,wuChargeDensityWave2022,liDiscoveryConjoinedCharge2022}, but surface-sensitive experiments repeatedly report additional modulations, most prominently a $4\times 1$ pattern in the Sb-terminated surface in CsV$_3$Sb$_5$ and RbV$_3$Sb$_5$~\cite{shumiyaIntrinsicNatureChiral2021,wangElectronicNatureChiral2021,liangThreeDimensionalChargeDensity2021,linUniaxialStrainTuning2024,zhaoCascadeCorrelatedElectron2021,chenRotonPairDensity2021,liUnidirectionalCoherentQuasiparticles2023,xingOpticalManipulationChargedensitywave2024,yeStructuralInstabilityCharge2022, nieChargedensitywavedrivenElectronicNematicity2022,liDiscoveryConjoinedCharge2022}. This $4\times 1$ modulation has remained controversial. It has been discussed as a surface-reconstruction effect~\cite{shumiyaIntrinsicNatureChiral2021,liDiscoveryConjoinedCharge2022,wangElectronicNatureChiral2021}, as a strain-selected variant~\cite{liangThreeDimensionalChargeDensity2021, linUniaxialStrainTuning2024}, or as an intrinsic instability competing with the bulk order~\cite{chenRotonPairDensity2021, zhaoCascadeCorrelatedElectron2021,yeStructuralInstabilityCharge2022}. This proliferation of plausible explanations underscores a broader challenge in these systems: with multiple near-degenerate ordering channels, even small perturbations can alter not only the transition temperature, but also the ordering wave vector and the underlying stabilization mechanism.

This sensitivity is ubiquitous across layered quantum materials, where tuning interlayer coupling and interlayer distance is now understood to be a powerful way to reorganize Fermi-surface instabilities and the competition between charge order and superconductivity~\cite{linEvidenceNestedQuasionedimensional2022, hsuAtomicallyresolvedInterlayerCharge2021,ugedaCharacterizationCollectiveGround2016,chenChargeDensityWave2015}.
The transition-metal dichalcogenides provide instructive benchmarks. In 2H-NbSe2, the $3\times 3$ charge order survives down to the monolayer limit, while superconductivity is strongly suppressed,
indicating that reducing interlayer coupling can decouple and reorder the balance between charge order formation and pairing~\cite{ugedaCharacterizationCollectiveGround2016}. In TiSe2, the charge order transition temperature increases in the single-layer limit, showing that interlayer coupling can be unfavorable to the charge order and
that the instability is not simply inherited from the bulk~\cite{chenChargeDensityWave2015}. Conversely, in 2H-TaS2, thinning suppresses the charge order while enhancing superconductivity, directly illustrating that the interlayer degree of freedom can tune competition between distinct Fermi-surface instabilities~\cite{yangEnhancedSuperconductivityWeakening2018}.
In 1T-TaS2, Raman and transport studies indicate distinct surface and bulk charge order behavior in ultrathin flakes, consistent with the emergence of layer-resolved ordering channels when interlayer interactions are weakened~\cite{heDistinctSurfaceBulk2016,tsenStructureControlCharge2015}. These examples indicate that dimensional crossover does not merely rescale energy scales; it can reorganize the hierarchy of instabilities and even change whether the lattice or the electronic subsystem plays the dominant role.

\begin{figure}[htbp]
    \centering
    \includegraphics[width=3.375in]{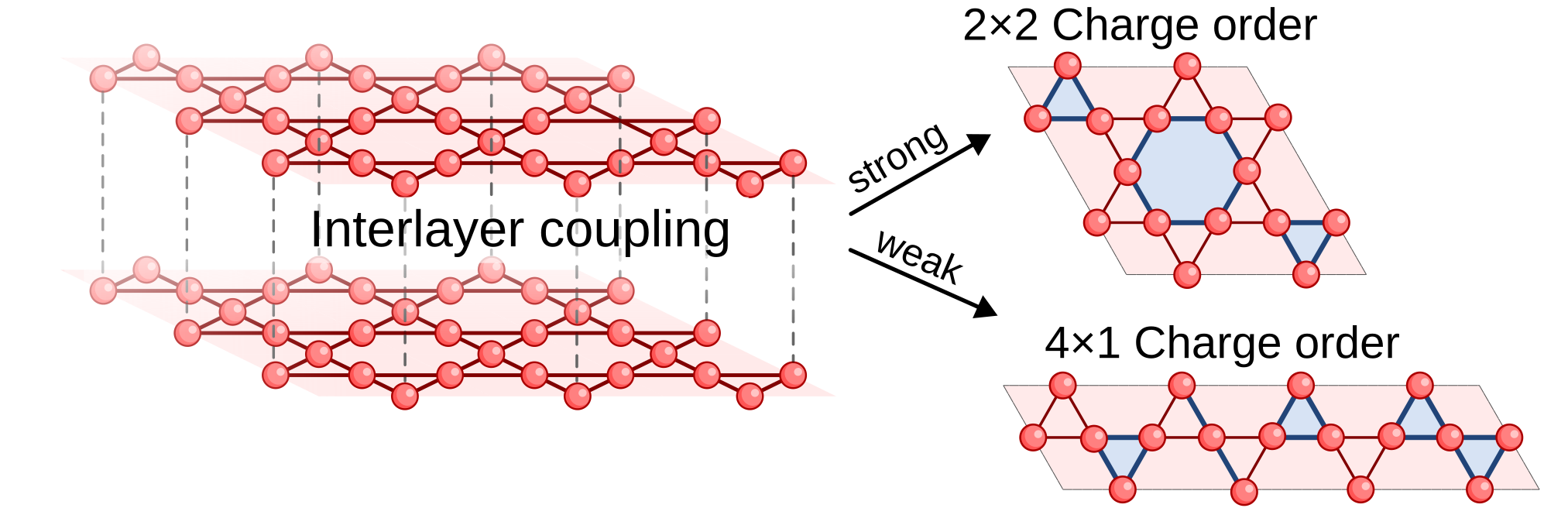}
    \caption{Schematic illustration of how interlayer coupling reconciles the competing $4 \times 1$ and $2 \times 2$ charge-order instabilities in CsV$_3$Sb$_5$.}
    \label{Figure1}
\end{figure}

This perspective motivates the central question of this work.  \textit{A}V$_3$Sb$_5$ (\textit{A} = K, Rb, Cs) family is quasi-two-dimensional but not strictly two-dimensional~\cite{ortizNewKagomePrototype2019,ortiz$mathrmCsmathrmV_3mathrmSb_5$$mathbbZ_2$Topological2020,liangThreeDimensionalChargeDensity2021,liObservationUnconventionalCharge2021}. The observation that bulk probes detect only $2 \times 2$ order, while surface measurements resolve both $2 \times 2$ and $4 \times 1$, points to interlayer coupling as a natural candidate for a microscopic selector of the $4 \times 1$ instability. Yet the literature has lacked a controlled, first-principles interpolation between the monolayer and bulk limits that isolates interlayer coupling as a single tuning knob.
Addressing this gap is essential for resolving the $4 \times 1$-order-controversy and, more broadly, for placing kagome charge orders within the general framework of dimensional selection of Fermi-surface instabilities. In this Letter, we provide such a conceptual advancement. Using density functional theory based phonon calculations together with electronic structure and
susceptibility analysis, we continuously tune the interlayer separation in CsV$_3$Sb$_5$ from the monolayer limit toward the bulk structure. We find that, in the monolayer, the lattice does not support the bulk-dominant $2 \times 2$ (M-point) instability, and the leading lattice instability occurs at $\text{q}=(1/4,0,0)$, consistent with a $4 \times 1$ modulation. As interlayer coupling
is introduced by reducing the interlayer distance, an M-point phonon progressively softens and becomes unstable already near $c \approx \qty{12.24}{\angstrom}$ for CsV$_3$Sb$_5$, evolving into the pronounced $2\times 2$ instability
at the bulk spacing. In this way, interlayer coupling provides the missing link that connects $4 \times 1$ and $2 \times 2$ tendencies within a unified microscopic picture (Fig.~\ref{Figure1}).

\begin{figure}[htbp]
    \centering
    \includegraphics[width=3.375in]{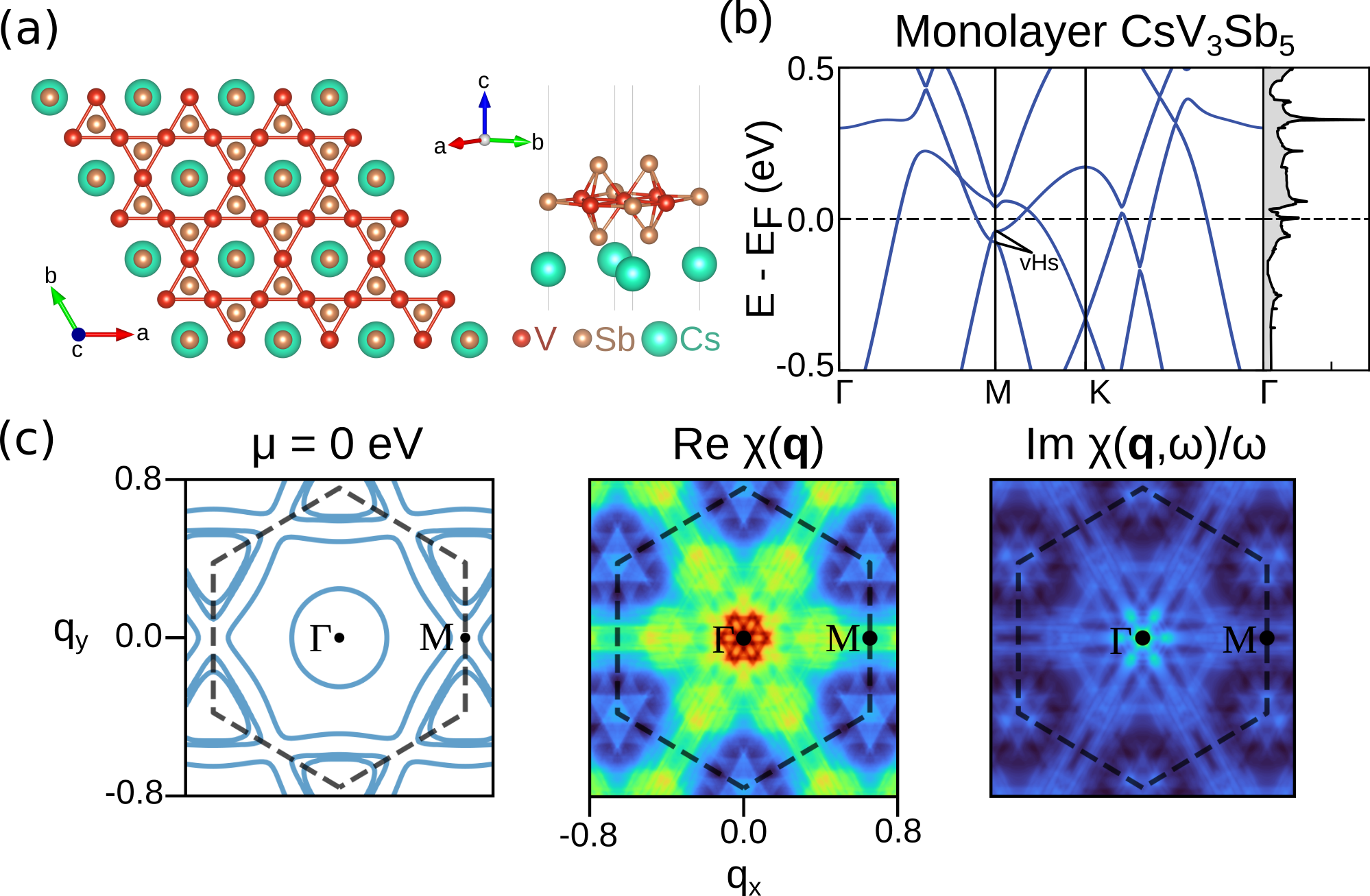}
    \caption{Electronic structure of CsV$_3$Sb$_5$ monolayer: (a) Crystal configuration, (b) band structure, (c) Fermi surface at zero chemical potential (left), real part (middle), and imaginary part (right) of electronic susceptibility.}
    \label{Figure2}
\end{figure}

We performed first-principles calculations within density functional theory using fully relaxed structures~\cite{kresseEfficiencyAbinitioTotal1996,kresseEfficientIterativeSchemes1996, supplemental}.
\nocite{blochlProjectorAugmentedwaveMethod1994,perdewGeneralizedGradientApproximation1996,grimmeConsistentAccurateInitio2010,blochlImprovedTetrahedronMethod1994,togoFirstPrinciplesPhonon2015,madsenBoltzTraP2ProgramInterpolating2018,pizziWannier90CommunityCode2020,damleCompressedRepresentationKohn2015, damleDisentanglementEntanglementUnified2018}
The monolayer limit is realized by introducing a large vacuum spacing (\qty{20}{\angstrom}) along the out-of-plane direction, thereby suppressing interlayer hybridization while
preserving the in-plane kagome structure (Fig.~\ref{Figure2}(a)). To interpolate between monolayer and bulk behavior, we systematically reduced the interlayer spacing while keeping the in-plane lattice constant fixed to the monolayer value, allowing us to study the role of out-of-plane coupling at fixed in-plane geometry. Phonon dispersions were computed using the non-diagonal finite difference method~\cite{supplemental, lloyd-williamsLatticeDynamicsElectronphonon2015}, and we evaluated the electronic band structure, density of states, Fermi surface, and susceptibility to diagnose electronic instabilities~\cite{wuChargeDensityWave2022}. This strategy directly tests whether electronic signatures of instability and lattice instabilities remain locked together as dimensionality changes.

We begin calculations in the monolayer limit, where the system is closest to the quasi two-dimensional Sb-terminated regime probed by surface-sensitive experiments (Fig.~\ref{Figure2}(a)). Unless otherwise noted, we focus on CsV$_3$Sb$_5$, which exhibits the largest energy lowering associated with charge order distortions among the kagome family \textit{A}V$_3$Sb$_5$ (\textit{A} = K, Rb, Cs). The electronic band structure in Fig.~\ref{Figure2}(b) (see also Figs.~S2 and S3 in the Supplemental Material~\cite{supplemental}) reveals vHs near the Fermi level at M point that motivated electronic mechanisms for $2 \times 2$ charge order formation in bulk \textit{A}V$_3$Sb$_5$ via Fermi surface nesting~\cite{dennerAnalysisChargeOrder2021}. However, susceptibility calculations show a ridge along the $\Gamma$-$M$ direction and a local minimum at M point in Fig.~\ref{Figure2}(c), suggesting that Fermi surface nesting does not produce a $2 \times 2$ charge order instability in monolayer CsV$_3$Sb$_5$~\cite{johannesFermiSurfaceNesting2008,wuChargeDensityWave2022,wangOriginChargeDensity2022}. The electronic structure, however, is not sufficient to explain $4\times 1$ charge order formation in CsV$_3$Sb$_5$ as observed via surface probes.

\begin{figure}[htbp]
    \centering
    \includegraphics[width=3.375in]{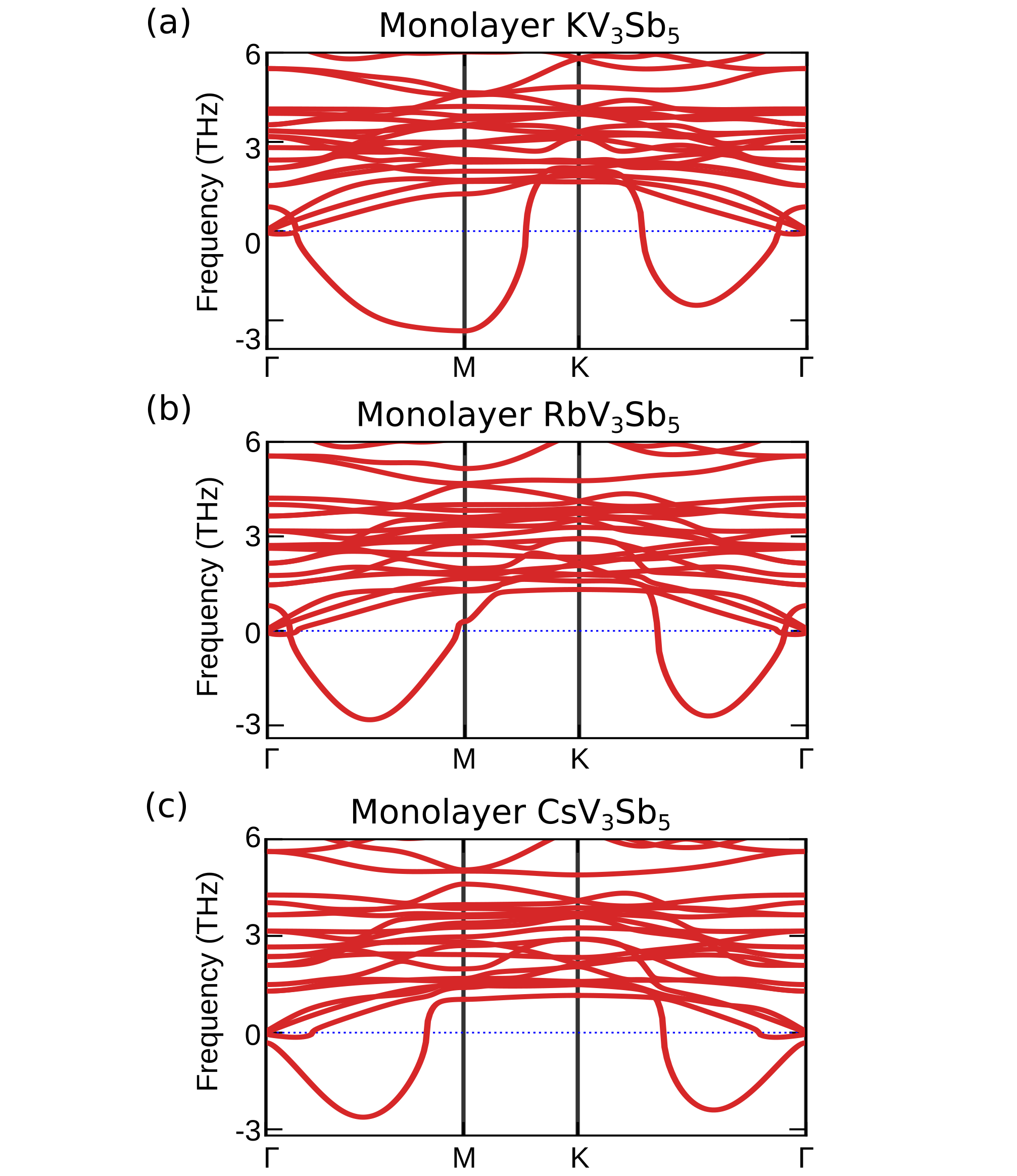}
    \caption{Phonon spectrum of monolayers: (a) KV$_3$Sb$_5$, (b) RbV$_3$Sb$_5$ and (c) CsV$_3$Sb$_5$. The M-point phonon mode is unstable exclusively in KV$_3$Sb$_5$.}
    \label{Figure3}
\end{figure}

In contrast, the phonon spectrum exhibits a distinct instability pattern in the kagome family in Fig.~\ref{Figure3}. KV$_3$Sb$_5$ monolayer shows imaginary phonons at $\text{M} =(1/2, 0, 0)$ or $2 \times 2$ lattice instability  as the bulk system in Fig.~\ref{Figure3}(a). However, in Fig.~\ref{Figure3}(b) and (c), the phonon branches of RbV$_3$Sb$_5$ and CsV$_3$Sb$_5$ monolayers do not become imaginary at the M point, indicating that the lattices do not support a $2 \times 2$ instability in the strict two-dimensional limit. Instead, the leading instability occurs at $\text{q} \approx (1/4, 0,0)$. A commensurate instability at $\text{q}=(m_1/n_1,m_2/n_2,m_3/n_3)$ folds to the $\Gamma$ point in an $n_1\times n_2\times n_3$ supercell Brillouin zone, producing a charge-ordered state of the corresponding periodicity. Because crystal symmetries generate a star of equivalent ordering vectors, the instability may condense either as a single-(Q) state or through the simultaneous condensation of multiple symmetry-related wave vectors, yielding a multi-(Q) charge-ordered phase \cite{mcmillanLandauTheoryChargedensity1975,mcmillanTheoryDiscommensurationsCommensurateincommensurate1976}. In monolayer \textit{A}V$_3$Sb$_5$ (\textit{A}=Cs, Rb), the instability at $\text{q}\approx(1/4,0,0)$ is consistent with the experimentally observed unidirectional $4\times1$ modulation, while in the presence of hexagonal symmetry, simultaneous condensation of the three symmetry-related ordering vectors would produce a $4\times4$ three-(Q) charge-ordered state. To understand the stability of these modulations with respect to the pristine structure, we modulate the pristine structure along the eigenmode of the instability with a commensurate supercell size. For the $1Q$ $4\times 4$ or $4\times 1$ charge order, we gain free energy of about $\qty{6.5}{meV/{f.u.}}$, showing the thermal stability of the modulated phase compared to the pristine phase at low temperature. This $1Q$ charge order modulation results in a reduction of $C_6$ to $C_2$ rotation symmetry. Nevertheless, our first-principles study suggests that the $3Q$ order is energetically more favorable than the $1Q$ order (Table~S1 in Supplemental Material~\cite{supplemental}). Accordingly, this $C_6$ to $C_2$ symmetry reduction in $1Q$ $4 \times 4$ order is a consequence of rotational symmetry breaking observed in experiments, not the other way around~\cite{liUnidirectionalCoherentQuasiparticles2023,zhaoCascadeCorrelatedElectron2021,fukushimaViolationEmergentRotational2024}.

However, in the CsV$_3$Sb$_5$ bilayer, the phonon spectrum shows instability at both M and $\text{q}=(1/4, 0, 0)$ points, indicating the coexistence of $2 \times 2$ and $4 \times 1$ instabilities in Fig.~S7(b) of the Supplemental Material~\cite{supplemental}. In addition, the M-point instability predominantly modulates the kagome plane confined between adjacent Cs layers. By contrast, the $4 \times 1$ instability is primarily confined to the Sb-terminated layer (Fig.~S7(c) and (d) in the Supplemental Material~\cite{supplemental}). To elucidate the origin of this coexistence, we computed the phonon spectrum of Cs$_2$V$_3$Sb$_5$  monolayer.
The phonon spectrum of Cs$_2$V$_3$Sb$_5$ monolayer shows a leading instability at M (Fig.~S8 in the Supplemental Material~\cite{supplemental}). This result should be viewed in accordance with the absence of $4\times 1$ modulation observed in Cs-terminated surface in STM measurements ~\cite{nieChargedensitywavedrivenElectronicNematicity2022}. In bilayer, the Sb-terminated kagome layer acts as monolayer CsV$_3$Sb$_5$ and the other layer as monolayer Cs$_2$V$_3$Sb$_5$. As a result, the phonon spectrum of bilayer exhibits both $4 \times 1$ and $2 \times 2$ instabilities (Fig.~S9 in the Supplemental Material~\cite{supplemental}). Given that the $4\times 1$ instability is absent in monolayer Cs$_2$V$_3$Sb$_5$, in contrast to the monolayer CsV$_3$Sb$_5$, we investigated additional surface stoichiometries in Sec.~S8 of the Supplemental Material~\cite{supplemental}. These findings illustrate that the emergence of charge order is critically dependent on surface stoichiometry (see Fig.~S12).

\begin{figure*}[htbp]
    \centering
    \includegraphics[width=\textwidth]{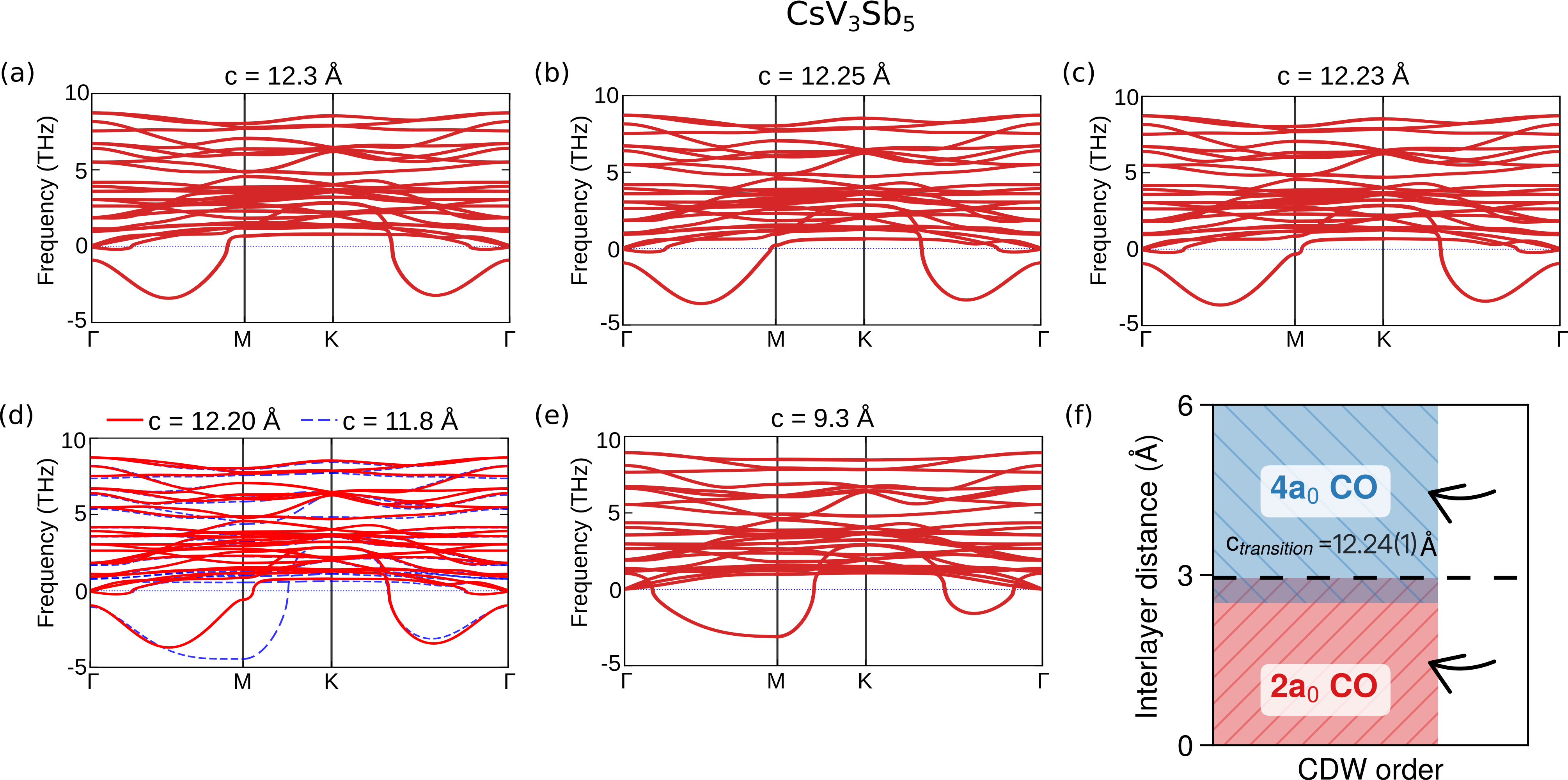}
    \caption{Tuning the interlayer coupling in  CsV$_3$Sb$_5$: (a--e) phonon spectra at varying interlayer spacing. The M-point phonon mode becomes unstable at $c=\qty{12.24 \pm 0.01}  {\angstrom}$ and the instability strengthens as the interlayer spacing approaches the bulk value $c=\qty{9.3}{\angstrom}$. (f) Evolution of the leading instabilities as a function of interlayer spacing, showing the emergence of $2 \times 2$ instability; both charge orders coexist in the shaded region. The interlayer distance is defined as the difference between the c-axis lattice constant and the bulk c-value}

    \label{Figure4}
\end{figure*}

Next, we tune interlayer coupling by reducing the interlayer separation from the monolayer limit toward the bulk value, relaxing the atomic coordinates at each step while keeping the in-plane lattice constant fixed to its monolayer value. This provides a direct and controlled route to unravel when and how the bulk-dominant $2 \times 2$ becomes operative. Strikingly, as the interlayer separation is reduced, the M-point phonon softens continuously (Fig.~\ref{Figure4}(a) and (b)). At about $c = \qty{12.24 \pm 0.01}{\angstrom}$, the M-point becomes imaginary, marking the onset of the $2 \times 2$ instability in Fig.~\ref{Figure4}(c) (see also Supplemental Material~\cite{supplemental}). As the interlayer spacing approaches the bulk value $c = \qty{9.30}{\angstrom}$~\cite{ortizNewKagomePrototype2019}, the imaginary
frequency deepens and the $2\times 2$ instability becomes dominant in Fig.~\ref{Figure4}(e). The critical value $\qty{12.24 \pm 0.01}{\angstrom}$ should be viewed as a coupling-strength threshold for the onset of $2\times2$ instability (see also Fig.~S13-S14) and, to the best of our knowledge, does not correspond to any currently reported experimental configuration. The bilayer result also complements this viewpoint: in bilayer CsV$_3$Sb$_5$, each layer has only one neighboring layer, in contrast to the bulk where each layer interacts with two neighboring layers. As a result, the interlayer coupling is weaker than in the bulk, leading the Sb-terminated layer to exhibit only $4\times1$ charge order~\cite{supplemental}. Therefore, the $4\times1$ instability arises at a physically realizable surface geometry with intrinsically weak interlayer coupling, requiring no artificial interlayer expansion.

Yet equally important is what does not happen in Fig.~\ref{Figure4}. The appearance of the M-point instability
does not trivially eliminate the $4 \times 1 $ tendency. Instead, the calculations reveal a landscape where multiple instabilities can coexist, with interlayer coupling selectively amplifying the $2\times 2$ channel until it alone survives (Fig.~\ref{Figure4}(c)-(e)). In this sense, interlayer coupling qualitatively reorganizes the hierarchy of ordering tendencies by activating a lattice-driven $2 \times 2$ route that is absent in the \emph{monolayer}. The key finding of our work is not simply that “interlayer coupling matters”; it is that interlayer coupling provides a microscopic selector that links two previously contested observations into a single coherent framework.

The monolayer result offers a concrete explanation for why $4 \times 1$ modulations appear so prominently in scanning tunneling microscopy: in the two-dimensional limit, the lattice instability landscape favors $4 \times 1$  while the $2 \times 2$ (M-point) lattice instability is absent (Fig.~\ref{Figure3}). The interpolation in Fig.~\ref{Figure4} then shows why bulk probes robustly observe $2 \times 2$: interlayer coupling
activates a soft M-point phonon and stabilizes $2 \times 2$ order as the layers approach the bulk spacing. This observation reveals that $4 \times 1$ and $2 \times 2$ are not unrelated anomalies. They are
intertwined outcomes of a single instability-prone energy landscape whose ranking is tuned by interlayer coupling (Fig.~\ref{Figure4}(f)).

This resolves the $4 \times 1$ controversy: one does not need to invoke a fundamentally different surface-only mechanism to account for $4\times1$ order. Instead, $4\times1$ order can be understood as an intrinsic two-dimensional instability (Fig.~S10-S11 in the Supplemental Material~\cite{supplemental}) that is revealed when interlayer coupling is reduced (Fig.~\ref{Figure4}). In that
sense, $4 \times 1$ charge order is a diagnostic of dimensional selection, not an extrinsic effect.

From the standpoint of microscopic physics, interlayer coupling can modify ordering in at least three intertwined ways~\cite{yangEnhancedSuperconductivityWeakening2018, ugedaCharacterizationCollectiveGround2016, chenChargeDensityWave2015, renLargeVariationInterlayer2025}. First, it changes electronic dispersion and three-dimensional warping, which can reshape susceptibility peaks and the phase space for scattering~\cite{yangEnhancedSuperconductivityWeakening2018, renLargeVariationInterlayer2025}. Second, it changes screening and the effective electron-phonon coupling~\cite{ugedaCharacterizationCollectiveGround2016,calandraEffectDimensionalityChargedensity2009}, both of which can be momentum dependent. Third, it modifies lattice dynamics directly, including the stiffness of
out-of-plane modes and their coupling to in-plane distortions~\cite{chenChargeDensityWave2015,duInterlayerEngineeringLattice2025}. The key outcome of our controlled interpolation is that these effects cooperate to selectively soften an M-point phonon and thereby activate the $2 \times 2$ lattice instability once interlayer coupling is sufficiently
strong. This conclusion should also be viewed in the broader context of recent works emphasizing that strong electron-phonon coupling and lattice anharmonicity can play a crucial role in stabilizing the charge order in \textit{A}V$_3$Sb$_5$~\cite{xieElectronphononCouplingCharge2022,youDiverseManifestationsElectronPhonon2025,gutierrez-amigoPhononCollapseAnharmonic2024,heAnharmonicStrongcouplingEffects2024}. Thus, our results identify a decisive structural control parameter that reorganizes the phonon instability landscape and thereby provides a direct route to reconcile apparently conflicting experimental observations in the literature.

The present results suggest concrete experimental tests. If $4 \times 1$ order is an intrinsic two-dimensional instability, then increasing the effective two-dimensionality should strengthen its visibility.
This can be pursued by exfoliation~\cite{songAnomalousEnhancementCharge2023}, by creating thin flakes under controlled strain environments~\cite{songCompetitionSuperconductivityCharge2021}, or by tuning interlayer coupling through intercalation, or substrate-induced spacing modifications~\cite{linUniaxialStrainTuning2024}. More generally, the results suggest that the multiplicity of charge order modulations in kagome metals may be an expected consequence of a nearby dimensional crossover rather than evidence for unrelated competing mechanisms.

In conclusion, we have provided a controlled first-principles dimensional understanding for \textit{A}V$_3$Sb$_5$ (\textit{A} = Rb, Cs) that identifies interlayer coupling as the microscopic selector at a fixed stoichiometry linking the controversial $4 \times 1$ modulation to the robust bulk $2 \times 2$ charge order. In the monolayer limit, the
lattice supports a leading instability at $\text{q}=(1/4, 0, 0)$ and exhibits no $2 \times 2$ (M-point) phonon instability. As interlayer coupling is introduced by reducing the interlayer separation, an M-point phonon progressively softens and becomes unstable near $c \approx \qty{12.24}{\angstrom}$, evolving into the strong bulk $2 \times 2$ instability at $c=\qty{9.3}{\angstrom}$. This establishes a unified framework for understanding why kagome metals display multiple charge-order wave vectors across experiments. More broadly, it illustrates a general principle for layered quantum materials: interlayer coupling can reorganize the hierarchy and even the mechanism of competing Fermi-surface instabilities, converting a two-dimensional instability landscape into a distinct three-dimensional ordered state.

\begin{acknowledgments}
This work used Anvil at Purdue University through allocation PHY260151 (awarded to MSH) from the Advanced Cyberinfrastructure Coordination Ecosystem: Services \& Support (ACCESS) program, which is supported by U.S. National Science Foundation grants \#2138259, \#2138286, \#2138307, \#2137603, and \#2138296.  MSH acknowledges support from the U.S. National Science Foundation award \#2610404 and the Samueli Foundation. The research conducted at Bangladesh University of Engineering and Technology (BUET) was partially supported by the BUET Basic Research Grant (Sub ID: 1111202106053).

\end{acknowledgments}

\section*{Data Availability Statement}
The data that support the findings of this study are available from the corresponding author upon reasonable request.

\bibliography{reference}
\end{document}